# Controlling Factors of $T_c$-Dome Structure in 1111-Type Iron Arsenide Superconductors

Satoru Matsuishi,[1,*] Takuya Maruyama,[2] Soshi Iimura[2] and Hideo Hosono[1,2,3]

[1]*Materials Research Center for Element Strategy, Tokyo Institute of Technology, 4259 Nagatsuta-cho, Midori-ku, Yokohama 226-8503, Japan*

[2]*Materials and Structures Laboratory, Tokyo Institute of Technology, 4259 Nagatsuta-cho, Midori-ku, Yokohama 226-8503, Japan*

[3]*Frontier Research Center, Tokyo Institute of Technology, 4259 Nagatsuta-cho, Midori-ku, Yokohama 226-8503, Japan*

**Abstract**

We investigated the effects of phosphorus substitution on the shape of the $T_c(x)$ dome in 1111-type SmFeAs$_{1-y}$P$_y$O$_{1-x}$H$_x$ ($0 < x < 0.5$). Hydride ion substitution of oxide sites (O$^{2-}$ → H$^-$) exerts a chemical pressure effect, i.e., a structural reduction of the $Pn$-Fe-$Pn$ angle $\alpha$ ($Pn$ = P, As) and also dopes electrons into the Fe$Pn$ layer to induce superconductivity. Isovalent phosphorus substitution (P$^{3-}$ → As$^{3-}$) can induce only a chemical pressure effect, i.e., an increase of $\alpha$ for La-substitution of Sm-sites. As $y$ increases from 0.0 to 0.5, the single $T_c$ dome gradually splits into two domes, similar to those of LaFeAsO$_{1-x}$H$_x$ with a $T_c$ valley at $x \approx 0.16$. We found that the $T_c$ valley is located around ($x$, $\alpha$) ≈ (0.16, 113°) for both SmFeAs$_{1-y}$P$_y$O$_{1-x}$H$_x$ and LaFeAsO$_{1-x}$H$_x$ series, irrespective of changes in the $Pn$ anion and $Ln$ cation species. This result suggests that suppression of $T_c$ leads to the emergence of a $T_c$ valley when both the shape of Fe$Pn_4$ tetrahedra represented by $\alpha$ and electron doping level of $x$ meet the above criterion in 1111 type iron oxypnictide superconductors.

*Corresponding author. E-mail: matsuishi.s.aa@m.titech.ac.jp

# I. INTRODUCTION

1111-type $Ln$FeAsO ($Ln$ = lanthanide) composed of an FeAs conducting layer and a $Ln$O insulating layer is a prototypical parent compound of iron-based superconductors[1–8] showing a tetragonal-orthorhombic structural transition at 120–160 K accompanied by antiferromagnetic ordering of Fe spins.[9–11] Superconductivity, with a critical temperature $T_c$ of up to 57.8 K,[12,13] emerges when the structural-magnetic transition is suppressed by electron doping of the FeAs-layer via fluoride ion (F$^-$) substitution of oxygen sites. The doped electron concentration $x$ vs temperature $T$ diagram of $Ln$FeAsO$_{1-x}$F$_x$ possess a superconducting (SC) region in the range $x$ = 0.1–0.2 adjacent to an antiferromagnetic (AFM) region at $x \approx 0.0$,[14] although the whole shape of the SC region has not been determined because the solubility limit of F$^-$ restricts $x$ to <0.2. The SC region adjacent to the AFM region is a common feature of the electronic phase diagram for many unconventional superconductors including iron pnictides.[15] Therefore, the electronic structure of $Ln$FeAsO$_{1-x}$F$_x$ has been well studied and the material dependence of $T_c$ on the spin-fluctuation-mediated pairing mechanism has been explained to some degree.[16–19]

Recently, we found that the SC region continues beyond $x \approx 0.4$ far from the AFM region by inducing high electron doping, with $x$ up to 0.5, using hydride ion (H$^-$) substitution in place of F$^-$.[20–23] In particular, LaFeAsO$_{1-x}$H$_x$ features a SC region with a unique shape which consists of a conventional dome (SC1) around $x \approx 0.1$ with a maximum $T_c$ ($T_c^{max}$) of 29 K and an additional dome (SC2) around $x \approx 0.3$ with a $T_c^{max}$ of 36 K. Inelastic neutron scattering measurements of LaFeAsO$_{1-x}$H$_x$ revealed a switch of momentum transfer of magnetic excitation from 1.14 Å$^{-1}$ for the first dome to 1.25–1.38 Å$^{-1}$ for the second dome, indicating that the pairing channels for SC1 and SC2 regions are different from each other.[24]

Here we focus on the effects of physical and chemical pressure on the $T_c$-dome shape reported in Ref. 22. When a physical pressure of up to 3 GPa was applied, the two SC regions for LaFeAsO$_{1-x}$H$_x$ merged into a single dome with $T_c^{max}$ = 47 K. Similar phenomenon was also induced by replacement of La with Ce, Sm and Gd. These results strongly suggest that the $T_c$ dome shape is controlled by the local structure of the FeAs$_4$ tetrahedron. As for the $Pn$-Fe-$Pn$ angle $\alpha$ of the Fe$Pn_4$ tetrahedron ($Pn$ = P, As), Lee $et\ al.$ have suggested that $T_c$ increases as $\alpha$ approaches that of a regular tetrahedron.[25] Further, Kuroki $et\ al.$ proposed on the basis of theoretical calculations that $T_c$ is raised as the $Pn$ height between the Fe plane $h_{Pn}$[18,19] is increased. However, because both structural parameters are related to each other by the equation; $h_{Pn} = r_{Fe-Pn}\cos\left(\frac{\alpha}{2}\right)$, where $r_{\text{Fe-}Pn}$ is the Fe-$Pn$ bond length, it is difficult to distinguish which parameters of $\alpha$ and $h_{Pn}$ has a dominant effect on the $T_c$-dome shape.

In this paper, we report the effects of phosphorus substitution on the $T_c(x)$ dome-shape of 1111-type SmFeAs$_{1-y}$P$_y$O$_{1-x}$H$_x$ ($0 \leq x < 0.5$). Isovalent P-substitution of the As site (P$^{3-}$→As$^{3-}$) induces a chemical pressure effect that changes $\alpha$ and $h_{Pn}$ in a similar manner to that induced by La-substitution of the Sm site. As a result, we found two-dome-shaped SC regions in $y = 0.45$ and $0.53$. Because P-substitution changes $r_{\text{Fe-}Pn}$, the resulting crystal structure can access the region in ($\alpha$, $h_{Pn}$) space unexplored by $Ln$-substitution. By considering the behavior of $\alpha$ and $h_{Pn}$ in SmFeAs$_{1-y}$P$_y$O$_{1-x}$H$_x$ and $Ln$FeAsO$_{1-x}$H$_x$ ($Ln$ = La, Ce, Sm) systems, we discuss the controlling factors for the $T_c(x)$ dome-shape of 1111-type Fe pnictides.

## II. EXPERIMENTAL

SmFeAs$_{1-y}$P$_y$O$_{1-x}$H$_x$ with nominal $y$ ($y_{\text{nom}}$) = 0.3, 0.5 and 0.6 were synthesized by the solid-state reaction of SmAs, FeAs, Fe$_2$As, Fe$_2$P, Sm$_2$O$_3$, and SmH$_2$ using a belt-type high pressure anvil cell. The metal arsenides and phosphides were prepared from their respective metals. SmH$_2$ was synthesized by heating metal samarium in a H$_2$ atmosphere. All starting materials and precursors for the synthesis were prepared in a glove box filled with purified Ar gas (H$_2$O, O$_2$ < 1 ppm). The mixture of stating materials was placed into a BN capsule with a mixture of Ca(OH)$_2$ and NaBH$_4$ as an excess hydrogen source, and then heated at 1473 K and 2 GPa for 30 min.

The amount of hydrogen incorporated into the resulting samples was evaluated by thermal desorption spectroscopy (TDS, ESCO TDS1200). Approximately 5 mg of sample were heated to 1373 K at a heating rate of 60 K/min under vacuum <10$^{-6}$ Pa. Hydrogen released from the sample, in the form of H$_2$ molecules, was ionized and detected by a quadrupole mass spectrometer as an ion with mass-to-charge ratio ($m/z$) = 2. Other non-hydrogen elemental compositions (Sm: Fe: As: P: O) were determined by an electron-probe micro-analyzer (EPMA, JEOL model JXA-8530F) equipped with a field-emission-type electron gun and wavelength dispersive x-ray detector. The micrometer-scale compositions within the main phase were probed on five to ten focal points, and the results were averaged.

Phase purity of the resulting samples was determined by powder x-ray diffraction (XRD) using a Bruker diffractometer model D8 ADVANCE (Cu rotating anode). The crystallographic parameters of the synthesized 1111-type compounds were determined by Rietveld analysis of XRD patterns using TOPAS code,[26] assuming a tetragonal symmetry (space group of $P4/nmmz$, lattice parameters $a = b \approx 0.39$ nm and $c \approx 0.84$ nm) with atomic positions of Sm (1/4, 1/4, $z_{Sm}$), Fe (3/4, 1/4, 1/2), Pn (1/4, 1/4, $z_{Pn}$) and O (3/4, 1/4, 0)). During the refinement of structural parameters, site occupancies of Pn and O sites were fixed to the values estimated by the EPMA.

Four-probe dc resistivity ($\rho$) and magnetic susceptibility ($\chi$) were measured in the temperature range 2–300 K, using a physical properties measurement system (PPMS) (Quantum Design, Inc.) with a vibrating sample magnetometer (VSM) attachment.

## III. RESULTS AND DISCUSSION

### A. P-substitution effect on $\alpha$ and $h_{Pn}$

Using high pressure techniques, we obtained polycrystalline pellets containing >80 weight % of the 1111-type phase of SmFeAs$_{1-y}$P$_y$O$_{1-x}$H$_x$ ($0 \leq x < 0.5$). The elemental composition analysis indicates that actual phosphorous content $y$ tended be less than the nominal content $y_{nom}$ of the starting mixtures, As a result, we obtained samples with $y$-values of $\approx 0.23$, $\approx 0.45$ and $\approx 0.53$ from starting mixtures with $y_{nom} = 0.3$, 0.5 and 0.6, respectively. The composition analysis is described in detail in the supplemental information. The amount of oxide vacancies measured almost matched the hydrogen content, indicating that the oxide ions are successfully replaced by hydride to form SmFeAs$_{1-y}$P$_y$O$_{1-x}$H$_x$. Previous neutron diffraction analysis and density functional theory calculations demonstrated that the hydrogen exclusively replaces O$^{2-}$ sites in the form of H$^-$ ions, and supplies an electron to the FePn layer to maintain charge neutrality.[21]

Figure 1(a) and (b) show the variations in the lattice parameters ($a$ and $c$) and $z$-coordinate of Sm and Pn sites ($z_{Sm}$ and $z_{Pn}$) as a function of $x$ in SmFeAs$_{1-y}$P$_y$O$_{1-x}$H$_x$ specified by $y = 0$, $\approx 0.23$, $\approx 0.45$ and $\approx 0.53$. With increases in $x$, the lattice parameters decrease and the $z$-coordinates increase, i.e., Sm$_4$O and FePn$_4$ tetrahedra become stretched along the $c$-axis upon compression of the crystal lattice. With increases in $y$, $a(x)$ and $c(x)$ curves are shifted down and the $z_{Sm}(x)$ curve is shifted up, indicating that the effects of P-substitution on these parameters are the same as those of H-substitution. In contrast, the effect of P-substitution on $z_{Pn}$ is the opposite to that of H-substitution. Because P-substitution effects are the same as those of La- and Ce-substitution of Sm sites, the ranges of the geometric parameters of FePn$_4$ tetrahedron, i.e., $\alpha$ and $h_{Pn}$ in SmFeAs$_{1-y}$P$_y$O$_{1-x}$H$_x$ partially overlap with those of the $Ln$FeAsO$_{1-x}$H$_x$ ($Ln$ = La–Sm) system.

Figure 2 shows $\alpha$ vs $h_{Pn}$ plots for $Ln$FeAsO$_{1-x}$H$_x$ and SmFeAs$_{1-y}$P$_y$O$_{1-x}$H$_x$. All the points of $Ln$FeAsO$_{1-x}$H are located within a narrow band like region corresponding to $r_{Fe-Pn}$ = 0.239–0.243 nm. In contrast to $Ln$- or H-substitution, which increases $r_{Fe-Pn}$ (+2 pm for 50% H-substitution and full substitution of La-to Sm), P-substitution decreases $r_{Fe-Pn}$ (−7 pm for 50% P-substitution) and expands the observable range in ($\alpha$, $h_{Pn}$) space to allow investigation of the structural dependence of the $T_c$-dome shape.

### B. P-substitution effect on $T_c$-dome shape

Figure 3 (a) shows the temperature dependence of resistivity for $y \approx 0.23$. At $x = 0$, a kink in resistivity due to structural/magnetic transitions was observed around 90 K. As $x$ increases, a sudden drop in resistivity to zero was observed at $x > 0.11$. The maximum onset $T_c$ was 44.5 K at $x = 0.20$ and this decreased to < 2 K at $x = 0.38$. For $y \approx 0.45$, as shown in Fig. 3(b), superconductivity was also observed at $x > 0.08$. The $T_c$ decreased from 22.5 K at $x = 0.08$ to 18 K at $x = 0.16$ but subsequently increased to 24 K at $x = 0.21$. Finally, $T_c$ decreased to < 2 K at $x = 0.43$. A similar drop in $T_c$ at $x = 0.16$ was also observed for $y \approx 0.53$ (See Fig. 3(c)). $T_c = 4$ K was observed at $x = 0.15$ between $x = 0.12$ and 0.18 showing $T_c^{max} = 14$ K.[27] It is notable that a drop of resistivity was observed at $\approx 5$ K in the $x = 0.0$ sample with $y \approx 0.53$. Because a drop in magnetic susceptibility was not observed in this temperature region, it is likely to originate from an AFM transition of Sm$^{3+}$ spins similar to that observed in SmFePO.[28] Behavior observed in $\chi$-$T$ profiles also reflected the development of bulk superconductivity as shown in Fig. 3(d)–(f). A drop in susceptibility due to superconducting transitions was observed at the magnetic onset $T_c^{mag}$ which was slightly lower than the resistivity onset $T_c$. As shown in Fig. 3(e) and (f), a drop of $T_c^{mag}$ was observed at $x = 0.15$–0.16 for $y \approx 0.45$ and 0.53. Figure 4 summarizes the magnetic onset $T_c^{mag}$ as a function of $x$. A $T_c^{mag}$ ($x$) curve for $y = 0$ taken from a previous paper[22] is also shown in this figure. In contrast to the single-dome-shaped SC region for $y = 0$ and $\approx 0.23$, a small valley is found at the center of the SC region in the $y \approx 0.45$ system. The split of the SC region was enhanced by further decrease in $y$ and the non-SC region ($T_c < 2$ K) was found at $x \approx 0.16$ in the $y \approx 0.53$ series.

### C. $T_c$-valley on $x$-$\alpha$ and $x$-$h_{Pn}$ spaces

Figure 5(a) is an $\alpha$ vs $x$ plot for SmFeAs$_{1-y}$P$_y$O$_{1-x}$H$_x$ and $Ln$FeAsO$_{1-x}$H$_x$. The symbol size of each point represents the $T_c$ value. Because of the chemical pressure effect of H-substitution, the $\alpha(x)$ curve of SmFeAsO$_{1-x}$H$_x$ has a negative slope and its intercept $\alpha_0$ at $x = 0$ increases with P- or La/Ce-substitution of Sm sites. Therefore, the ranges of $\alpha(x)$ curves for $y = 0.23$–$0.53$ overlap with those for $Ln$ = La, Ce. The regions of the $T_c$-valleys for $y = 0.45$ and $0.53$ are denoted by a light blue area and those for $Ln$ = La by a light pink area. Both regions are clearly located at the same area around $(x, \alpha) = (\approx 0.16, \approx 113°)$. In other words, two-dome shaped $T_c(x)$ curves are observed when the $\alpha(x)$ curve has a relatively large $\alpha_0$ (>114°) and passes through the region around ($\approx 0.16, \approx 113°$). Figure 5(b) is an analogous $x$ vs $h_{Pn}$ plot. In $x$-$h_{Pn}$ space, the $T_c$ valley for $y = 0.45$–$0.53$ is located at a position with different $h_{Pn}$ (0.130 nm for $y = 0.45$–$0.53$; 0.134 nm for $Ln$ = La). In addition, a drop of $T_c$ is not observed in the $y \approx 0.43$ series, while their $h_{Pn}(x)$ curve crosses the region of the $T_c$-valley of LaFeAsO$_{1-x}$H$_x$. These results indicate that the emergence of a $T_c$-valley depends on $\alpha$, and not $h_{Pn}$, irrespective of the $Pn$ anion and $Ln$ cation species.

Finally, we discuss the effects of $\alpha$ on $T_c$. Because of the relationship between $\alpha$ and $T_c$ in $Ln$Fe$Pn$O$_{1-x}$H$_x$, the "pairing glue" that induces superconductivity appears to depend on $\alpha$. A candidate for this glue is the electron-phonon interaction associated with $\alpha$. Saito *et al.* suggested in their orbital fluctuation theory that electron-phonon interactions are strongest when the FeAs$_4$ tetrahedron has a regular shape ($\alpha = 109.5°$).[29] This theory can empirically explain the maximum $T_c$ occurring when $\alpha$ is 109.5°. However, in comparing $\alpha$ and $x$ dependence of $T_c$, the resulting $T_c$-$\alpha$ relation is not monotonic, but does depend on $x$. Another possibility is spin fluctuation enhanced by hopping between Fe $3d$ orbitals via As $4p$ orbitals. The hopping integral, i.e., the overlap integral of Fe $3d$ and As $4p$ orbital is directly determined by Fe-As-Fe angle $\gamma$ having a one-to-one relation with $\alpha$: $\sin(\gamma/2) = 1/\sqrt{2} \cdot \sin(\alpha/2)$. Very recently, a calculation using the fluctuation exchange approximation successfully reproduced the shape-variation of $T_c$-$x$ with $\alpha$ in a $Ln$FeAsO$_{1-x}$H$_x$ system; from a two-dome shape at large $\alpha$ to one-dome shape at small $\alpha$.[30] The primary suggestion by this calculation is that the spin-fluctuation enhancement mechanism is different between low-$x$ and high-$x$ regions. The spin fluctuation in the former region is caused by nesting of a Fermi surface composed of Fe $3d_{xz,yz}$ and $3d_{xy}$ orbitals, while that of the latter originates from real space motion of electrons within the $3d_{xy}$ orbitals, i.e., next nearest neighbor diagonal hopping ($t_2$) dominates the nearest neighbor hopping ($t_1$). In this scheme, $\alpha$ plays an important role in determining the magnitude of spin fluctuations and $T_c$ because both $t_2$ and indirect $t_1$

($t_1^{indirect}$) are controlled by hopping via As 4$p$ orbitals. Given the experimental findings and the theoretical real-space picture of electronic spin, $\alpha$ is the decisive factor for the shape-variation of $T_c$-$x$ in $Ln$FeAsO$_{1-x}$H$_x$ systems.

IV. **SUMMARY**

We found a two-dome-shaped SC region in the $x$-$T$ diagram of SmFeAs$_{1-y}$P$_y$O$_{1-x}$H$_x$ with $y$ = 0.45 and 0.53. This finding agrees with previous results on LaFeAsO$_{1-x}$H$_x$. By investigating the correlation between $T_c$ and structural parameters of the Fe$Pn_4$ tetrahedron, we confirmed that the $T_c$-valley separating two the SC domes is located at $x \approx 0.16$ and $\alpha \approx 113°$ in both SmFeAs$_{1-y}$P$_y$O$_{1-x}$H$_x$ and LaFeAsO$_{1-x}$H$_x$ systems. These results indicate that the emergence of two SC domes is generally induced in 1111-type iron oxypnictides when the electron doping level matches the shape of the Fe$Pn_4$ tetrahedron

**Acknowledgement**

This work was supported by the JSPS FIRST project and the MEXT Element Strategy Initiative project.

**Figures Captions**

**Figure 1** (Color online) Crystallographic parameters of SmFeAs$_{1-y}$P$_y$O$_{1-x}$H$_x$ with $y = 0, \approx 0.23, \approx 0.45$ and $\approx 0.53$ as a function of $x$. (a) Lattice parameters $a$ and $c$. (b) $z$-coordinate of Sm and $Pn$ sites ($z_{Sm}$ and $z_{Pn}$). Parameters for $y = 0$ are taken from previous literature.[20]

**Figure 2** (Color online) $\alpha$ vs $h_{Pn}$ plots of $Ln$FeAsO$_{1-x}$H$_x$ ($Ln$ = La, Ce and Sm)[22] and SmFeAs$_{1-y}$P$_y$O$_{1-x}$H$_x$ ($y = 0.23, 0.45$ and $0.53$). Lines are guides to the eye tracing the $\alpha$-$h_{Pn}$ relation for specified $y$ and $Ln$. In $Ln$FeAsO$_{1-x}$H$_x$, the change in Fe-As bond length is small ($r_{Fe-Pn} = 0.239$–$0.243$ nm) and $h_{Pn}$ is strongly linked to $\alpha$. By replacing As by P, $r_{Fe-Pn}$ clearly decreases.

**Figure 3** (Color online) Temperature dependence of electrical resistivity (a–c) and magnetic susceptibility (b–f) in SmFeAs$_{1-y}$P$_y$O$_{1-x}$H$_x$ with $y \approx 0.23, \approx 0.45$ and $\approx 0.53$.

**Figure 4** (Color online) Magnetic onset $T_c$ of SmFeAs$_{1-y}$P$_y$O$_{1-x}$H$_x$ as a function of $x$ with $y = 0, \approx 0.23, \approx 0.45$ and $\approx 0.53$. $T_c$ values of $y = 0$ samples were taken from reference 20.

**Figure 5** (Color online) Correlation between the emergence of the $T_c$ valley and structural or doping parameters in SmFeAs$_{1-y}$P$_y$O$_{1-x}$H$_x$ ($y = 0, \approx 0.23, \approx 0.45$ and $\approx 0.53$; blue solid symbols) and $Ln$FeAsO$_{1-x}$H$_x$ ($Ln$ = La and Ce); red empty symbols). (a) $\alpha$ angle of Fe$Pn_4$ tetrahedron as a function of $x$. Lines are guides to the eye tracing $\alpha(x)$ for specified $y$ and $Ln$. The symbol size indicates the $T_c$ value. The region of the $T_c$ valley for SmFeAs$_{1-y}$P$_y$O$_{1-x}$H$_x$ is denoted by a light blue area around $(x, \alpha) = (0.16, 113)$ which overlaps with that of LaFeAsO$_{1-x}$H$_x$ denoted by a light pink colored area. (b) $h_{Pn}$ vs $x$ plot. The regions of the $T_c$ valleys in each system do not overlap.

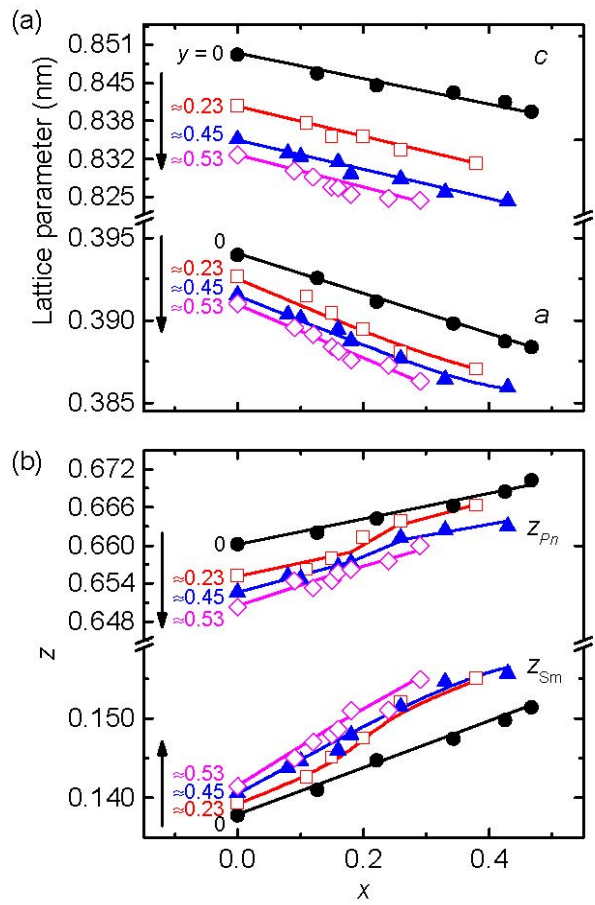

Fig.1

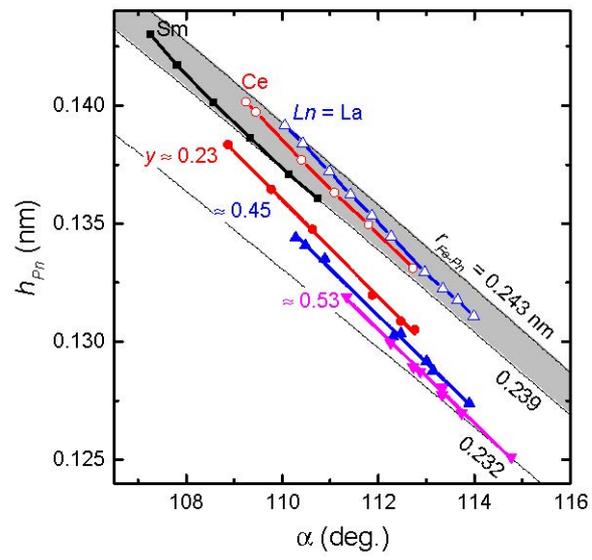

Fig.2

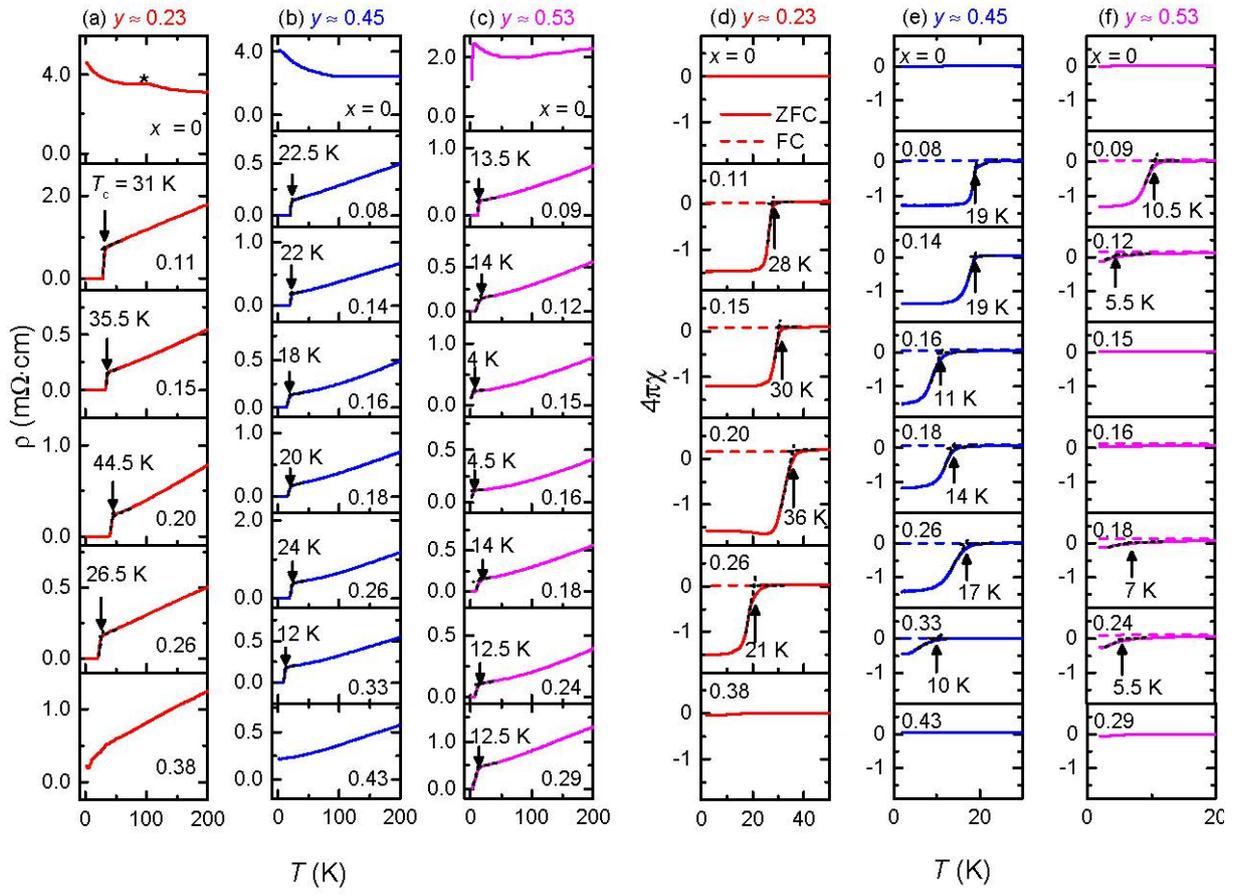

Fig.3

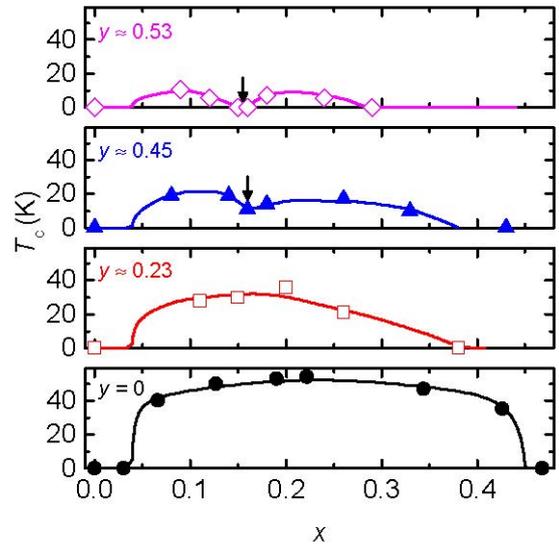

Fig.4

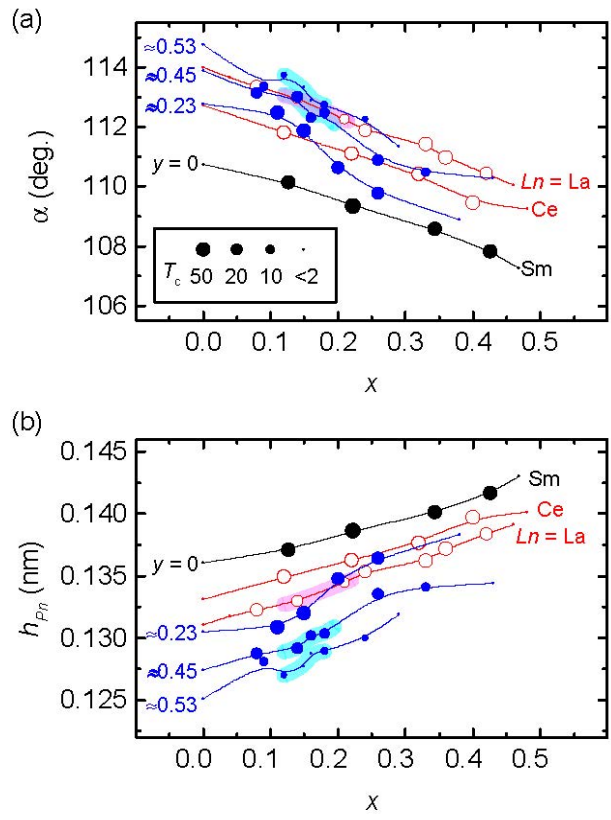

Fig.5

## Supplemental Information

**Chemical Compositions**

Figure S1(a)–(c) show $x$ and $y$ values analyzed in the synthesized SmFeAs$_{1-y}$P$_y$O$_{1-x}$H$_x$ samples with the $y_{nom}$ value specified as a function of nominal $x$ ($x_{nom}$). Because of segregation of impurity phases such as Fe$_2$P, the phosphorus content measured by EPMA was prone to be less than $y_{nom}$. For $x_{nom} > 0$ ($y_{EPMA}$; ≈ 0.23, ≈ 0.45 and ≈ 0.53 in $y_{nom}$ = 0.3, 0.5 and 0.6 samples, respectively), the amount of oxygen vacancies estimated by EPMA ($x_{EPMA}$) is proportional to $x_{nom}$ and agrees well with the hydrogen content measured by thermal desorption spectroscopy ($x_{TDS}$), indicating that oxide is successfully replaced by hydride. Previous neutron diffraction analysis and density functional theory calculations demonstrated that hydrogen exclusively replaces O$^{2-}$ sites in the form of H$^-$ supplying an electron to the Fe$Pn$ layer to maintain charge neutrality.[1] In contrast, for the $x_{nom}$ = 0 samples, the EPMA analysis indicated an oxygen deficiency of up to ≈ 10%. According to the previous papers, such a large amount of oxygen vacancies would typically induce superconductivity.[2–4] However, no superconductivity was observed in these samples, indicating a low concentration of oxygen vacancies. A plausible reason for the observed oxygen deficiency is the incorporation of phosphide ions (P$^{3-}$) into oxygen sites. In the samples with specified $y_{nom}$, the $x_{nom}$ = 0 samples shows a relatively high $y$ value. Therefore, in the text, $x$ and $y$ are expressed by $x_{EPMA}$ and $y_{EPMA}$, respectively, except for $x$ = 0, which is indicated by $x_{nom}$ = 0.0.

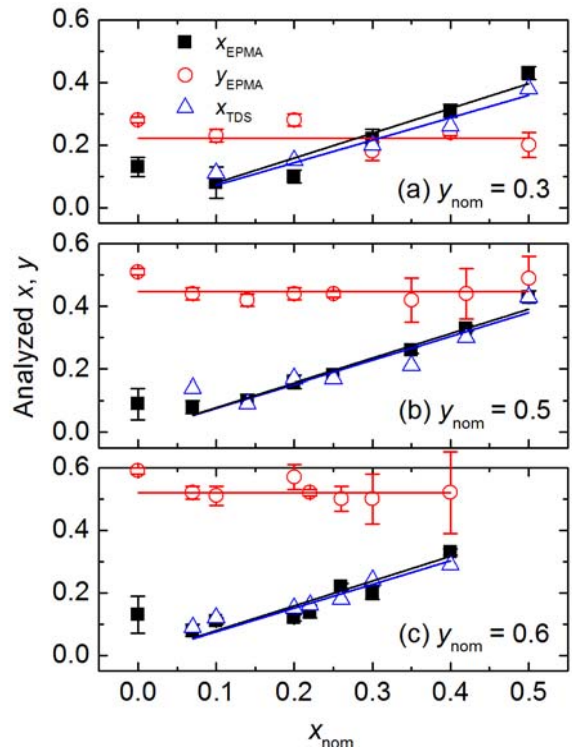

**Figure S1** Analyzed $x$ and $y$ values in SmFeAs$_{1-y}$P$_y$O$_{1-x}$H$_x$ as a function of nominal $x$ ($x_{nom}$). (a) $y_{nom}$ = 0.3, (b) $y_{nom}$ = 0.5 and (c) $y_{nom}$ = 0.6.